\documentclass[twocolumn, a4paper]{article}
\usepackage{0_arxiv}

\usepackage[utf8]{inputenc} 
\usepackage[T1]{fontenc}    
\usepackage{hyperref}       
\usepackage{url}            
\usepackage{booktabs}       
\usepackage{amsfonts}       
\usepackage{nicefrac}       
\usepackage{microtype}      
\usepackage{fancyhdr}       
\usepackage{lipsum}
\usepackage[ddmmyyyy]{datetime}
\newdateformat{mydate}{\twodigit{\THEDAY}{ }\shortmontfsubfihname[\THEMONTH], \THEYEAR}

\usepackage{amsmath} 
\usepackage{amssymb} 
\usepackage{dsfont}
\usepackage[pdftex,dvipsnames]{xcolor}
\usepackage[colorinlistoftodos,prependcaption]{todonotes}

\usepackage[babel,german=guillemets]{csquotes}
\usepackage{float}
\usepackage{graphicx} 
\usepackage{listings}
\usepackage{algorithm}
\usepackage[noend]{algpseudocode}
\usepackage{makecell}
\usepackage{microtype} 
\usepackage{mparhack}
\usepackage{multicol} 
\usepackage{multirow} 
\usepackage{mathtools}
\usepackage{paralist} 
\usepackage{wrapfig} 
\usepackage[verbose]{placeins}
\usepackage[format=plain, font={small,up}, labelfont=bf, justification=justified, labelsep=period, skip=10pt]{caption}

\usepackage{subcaption}
\usepackage[nottoc]{tocbibind}
\usepackage{my_newcommands}
\usepackage{algpseudocode}
\usepackage{algorithm}
\usepackage{cooltooltips}
\usepackage{wrapfig}

\title{$d$CG - differentiable connected geometries for AI-compatible multi-domain optimization and inverse design}

\author{
    Alexander Luce \\ 
    University Erlangen-Nürnberg \&\\
    Max Planck Institute \\
    for the Science of Light\\
    Erlangen \& \\
    ams OSRAM\\
    Regensburg \\
\And
    Daniel Grünbaum \\ 
    ams OSRAM\\
    Regensburg\\
\And
   Florian Marquardt \\ 
   University Erlangen-Nürnberg \&\\
   Max Planck Institute \\
   for the Science of Light\\
   Erlangen \\
}

\begin{document}

\twocolumn[ 
  \begin{@twocolumnfalse} 
  
\maketitle

\begin{abstract}
In the domain of geometry and topology optimization, discovering geometries that optimally satisfy specific problem criteria is a complex challenge in both engineering and scientific research. In this work, we propose a new approach for the creation of multidomain connected geometries that are designed to work with automatic differentiation. We introduce the concept of differentiable Connected Geometries (dCG), discussing its theoretical aspects and illustrating its application through simple toy examples and a more sophisticated photonic optimization task. Since these geometries are built upon the principles of automatic differentiation, they are compatible with existing deep learning frameworks, a feature we demonstrate via the application examples. This methodology provides a systematic way to approach geometric design and optimization in computational fields involving dependent geometries, potentially improving the efficiency and effectiveness of optimization tasks in scientific and engineering applications.
\end{abstract}
\vspace{0.35cm}

  \end{@twocolumnfalse} 
]


\section{Introduction}
\label{sec:1_introduction}
The field of shape and topology optimization \cite{Sigmund:2013:Topology_optimization_approaches, Guo:2010:Recent_development_in_structural_design_and_optimization} is concerned with finding optimal shapes that achieve a certain physical goal, such as having an optimal shape for an aircraft wing \cite{Schramm:2018:Optimization_of_Airfoils_Using_the_Adjoint_Approach_and_the_Influence_of_Adjoint_Turbulent_Viscosity}, the design for an optimal heat exchanger \cite{Kametani:2023:Adjoint_Sensitivity_Analysis_for_Heat_Transfer_Enhancements_in_Structured_Channels, Morimoto:2010:Optimal_Shape_Design_of_Compact_Heat_Exchangers_Based_on_Adjoint_Analysis_of_Momentum_and_Heat_Transfer}, a waveguide profile or photonic crystals for integrated photonic circuits (PIC) \cite{Molesky:2018:Inverse_design_in_nanophotonics, Wang:2017:Optimization_of_photonic_crystal_cavities, Jensen:2011:Topology_optimization_for_nano-photonics, Lalau:2013:Adjoint_shape_optimization_applied_to_electromagnetic_design}. Since typical geometrical problems have a huge amount of parameters, gradient based optimization is typically required for shape and topology optimization \cite{Keogh:2017:Curse_of_Dimensionality, Lebbe:2019:Contribution_in_topological_optimization_and_application_to_nanophotonics, Luce:2024:Merging_automatic_differentiation_and_the_adjoint_method_for_photonic_inverse_design}. The process is as follows, the optimization is started with an initial guess for a design. The design is simulated by solving the underlying physical equations with an appropriate solver and a loss is determined. To be able to determine how the design must be changed in order for the loss to decrease, we need to compute the gradients of the design parameters. For a numerical simulation, this can be done efficiently via the adjoint method \cite{Johnson:2021:Notes_on_Adjoint_Methods}. This involves solving a second system of equation, the so called adjoint system. From the two solutions, it is possible to determine the gradients on the design geometry \cite{Lebbe:2019:Contribution_in_topological_optimization_and_application_to_nanophotonics,  Delfour:1991:Velocit_Method_and_Lagrangian_Formulation_for_the_Computation_of_the_Shape_Hessian, Delfour:2011:Shapes_and_Geometries}.

By following a gradient descent scheme \cite{Nesterov:2018:Lectures_on_Convex_Optimization, Fletcher:1987:Practical_Methods_of_Optimization} the shape is deformed to decrease a given loss function. Typically, the gradients are computed on the surface of the designable object. Often, a certain component, ie. the waveguide of a PIC \cite{Molesky:2018:Inverse_design_in_nanophotonics}, is embedded in an ambient medium - a clear distinction is made between design region and surrounding medium. The optimization then manipulates the design domain within the ambient domain following the gradient descent scheme. This entails that designing a multi-component device that consists of multiple distinct domains with different boundaries and physical properties is difficult since nothing prevents domains from overlapping each other, creating ambiguous domains. 

To make it possible to incorporate multiple domains with different material properties into the same optimization region different techniques were developed. The approach by Tajs-Zielinka et al. \cite{Tajs-Zielinka:2021:Multi-Domain_and_Multi-Material_Topology_Optimization_in_Design_and_Strengthening_of_Innovative_Sustainable_Structures} incorporated multiple domains with different materials but fixed boundaries which can lead to suboptimal designs since the chosen domain geometry plays a significant role. Closely related are moving morphable components (MMC) \cite{Zhang:2018:Topology_optimization_with_multiple_materials_via_moving_morphable_component_(MMC)_method, Guo:2014:Doing_Topology_Optimization_Explicitly_and_Geometrically-A_New_Moving_Morphable_Components_Based_Framework}. MMC is a framework for topology optimization that allows the creation of complex structures by combining simple components. The components are designed to be easily modifiable and can be moved, resized, or deleted to create new structures. Often, the optimization domain is filled with a set of predefined topology description functions (TDF) \cite{de_Ruiter:2004:Topology_optimization_using_a_topology_description_function}. TDFs are mathematical functions that describe the shape and layout of a structure. The geometry components represented by the TDFs can be moved around, overlapped, and reshaped according to the parametrization. Some of the most common types include density-based functions, level set functions \cite{Wang:2004:Level-set_method_for_design_of_multi-phase_elastic_and_thermoelastic_materials}, and phase field functions \cite{Osher:2003:Level_set_methods_and_dynamic_implicit_surfaces}.

On top of that, there is usually a functional relationship between different components of a physical problem, ie. modifying the taper of a waveguide influences the beam profile which in turn requires a different mode-splitter design which might further influence components down the line. As a consequence components need to be designed one after another which is prolonging the development and requires additional effort. Further, great care needs to be taken that the components further down the line have no influence on the behavior of the previous components and thus each component works independently. Otherwise, the optimization for each component needs to be performed again to take the change of the previous optimization into account. 

Following the developments of machine- and reinforcement learning, these advancements were applied to topology and shape optimization. Here, the difficulty is to couple the neural networks to the domain and geometry description. In many cases, the geometry is interpreted pixelwise and discrete information about the geometry is given to the network such as a generative adversarial network (GAN) \cite{Nie:2021:TopologyGAN:_Topology_Optimization_Using_Generative_Adversarial_Networks_Based_on_Physical_Fields_Over_the_Initial_Domain, 2021:Ma:Deep_learning_for_the_design_of_photonic_structures}. In the work of Lei et al \cite{Lei:2018:Machine_Learning_Driven_Real_Time_Topology_Optimization}, they trained a non-linear regression model to propose an optimal set of TDFs for an MMC approach. In both cases, the neural networks only learn the dynamics implicitly by data learning on a dataset. If the optimal solution of the problem under investigation is outside of the dataset distribution, the neural network might never propose the actual optimum. Additionally, the approach is inflexible since a new dataset must be created when the problem is only reformulated slightly.


This work presents an approach to connect automatic gradient computation \cite{Leal:2023:autodiff, Baydin:2017:Automatic_Differentiation_in_Machine_Learning:_A_Survey} with the geometry definition for a physical optimization problem. Our framework proposes connected regions of space, and we explain how gradients are computed on the appropriate geometry boundaries to allow integration with existing numerical adjoint solvers and shape optimization techniques. By connecting geometries via differentiable functions, we take the sensitivity of changing the parent geometry into account, which enables relative definitions to connect geometries. This approach is particularly useful for advanced functionality, where a complicated interaction between individual components is expected. Relative geometries are often easier to constrain and enable better control over the generated minimum. With this approach, we hope to provide a more intuitive and efficient way to solve physical optimization problems. Additionally, computing gradients directly on the geometry enables neural networks and reinforcement learning agents to easily integrate with the environment and obtain gradient information from the geometry itself directly within the same autodifferentiation framework.

\section{Differentiable and connected geometries}
\label{sec:connected_geometries}
General individual geometries within an ambient medium can be deformed on their entire surface. Any point on the surface of a geometry can displaced and the shape of the geometry warped. For connected geometries, this is not necessarily the case anymore since any geometries can overlap. Consequently, the geometries must be ordered to establish which are the parent and child components. 

\begin{figure}[ht]
  \centering
  \includegraphics[angle=0, trim = 0cm 0cm 0cm 0cm, clip, width = \columnwidth]{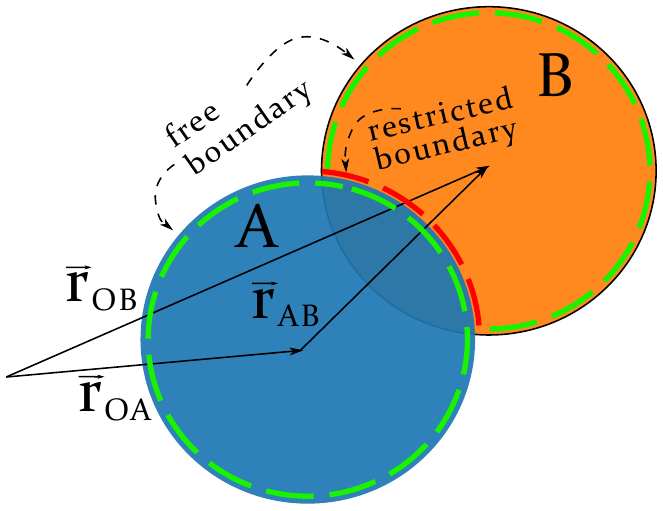}
  \caption{Two simple connected geometries. $A$ is the root geometry and parent of $B$. The entire boundary of $A$ is free while the boundary of $B$ at the intersection with with its parent is restricted. The restricted boundary can't be modified by changing parameters of $B$. \label{fig:atomic_geometries}}
\end{figure}

Consider two different geometries, for the sake of simplicity, assume two circles which are partially overlapping. Without any ordering, it is not clear which of the two domains will be realized in a physical setting. We must make a choice which of the two geometries has priority and is therefore the parent. Let's denote one of the geometries by $A = \{\vec{x} | x^2 + y^2 - r_A \leq \vec{r}_A\} $ which is the set of all points within a circle. $\vec{r}_A$ denotes the displacement of the circle from the origin. The other is denoted by a second set $B = \{ \vec{x} | x^2 + y^2 - r_B \leq \vec{r}_B \}$. Let's impose an ordering on the geometries where $A$ becomes the parent of $B$. This implies that $B$ is now reduced to $B / A$ which is given by $B=\{ \vec{x} | x^2 + y^2 - r_B \leq \vec{r}_B \wedge \vec{x} \notin A \}$. Now, since $B$ is partially defined by $A$, part of the surface of $B$ is not free anymore, it depends on the geometry of $A$. This geometry surface is therefore denoted as \textit{restricted boundary} in contrast to any part of the boundary which is defined directly by the parameters is called \textit{free boundary}.

The distinction is important, any gradients on the restricted boundary of $B$ do not influence the parameters of $B$ directly, only gradients on the free boundary of $B$ need to be taken into account. The restricted boundary of $B$ is, by definition, a free boundary on $A$, see \autoref{fig:atomic_geometries}. 

Formally, we can write:\\
Def: a free boundary, denoted by $\p_f$:
\begin{align}
    \bigl\{\p_f A_i \subseteq \partial A_i |\, & A_i(p'),\, p' = p_0 + \varepsilon \\&\Rightarrow \p A_i(p') - \p A_i(p) \neq 0\bigr\}. 
\end{align}
The free boundary indicates a completely free boundary where changes to any geometric parameters $p$ would result in a change of the subset of the boundary. Thus, the gradients on that boundary needs to be backpropagated when calculating the gradients for the parameter $p$. 
For a restricted boundary, changes to a parameter do not influence the boundary, hence:\\
Def: a restricted boundary, denoted by $\p_r$:
\begin{align}
    \bigl\{\p_r A_i \subseteq \partial A_i |\, & A_i(p'),\, p' = p_0 + \varepsilon \\&\Rightarrow \p A_i(p') - \p A_i(p) = 0\bigr\}. 
\end{align}
denotes a part of the boundary which is independent from change to geometric parameters. 

The free boundary is the complement of the restricted boundary: 
\begin{align}
    \p_f A_i = \partial A_i / \partial_r A_i,
\end{align}
thus any part of the boundary $\p A_i$ is either free or restricted. A geometry without any parent can't have a restricted boundary and therefore $\p_f A_i = \partial A_i$, applies.

\begin{figure}[ht]
  \centering
  \includegraphics[angle=0, trim = 0cm 0cm 0cm 0cm, clip, width = \columnwidth]{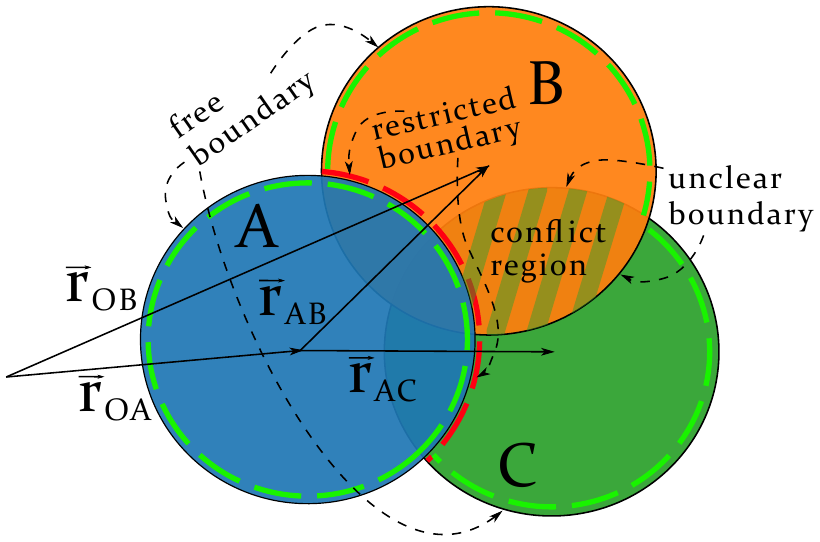}
  \caption{Conflict arising from unclear ordering of the child geometries of $A$. It is unclear which geometry is realized at the intersection of $B$ and $C$. We need to impose an ordering to resolve the conflict. The conflict is resolved by selecting $B$ as the parent of $C$. The conflict region becomes part of $B$. As a consequence, only the lower boundary exists, it becomes a free boundary for $B$ and a restricted boundary for $C$.\label{fig:atomic_geometries_conflict}}
\end{figure}

The ordering also implies a functional relationship between the geometries. Although it is possible to reference a child geometry on the origin, it is typically desired to move the child relative to the parent and make it dependent on some geometrical feature of the parent. In the example shown in \autoref{fig:atomic_geometries}, the center of the child $B$ is referenced on the center of the parent $A$ by a vector $\Vec{r}_{AB}$. This however entails a problem for optimization. Changing the absolute position of the parent geometry has a direct influence on the child due to the relative definition. This displacement can have adverse effects if the behavior of the child is not taken into account appropriately \footnote{Imagine a situation where moving $A$ up decreases the loss slightly but moving $B$ up increases the loss strongly. Since $B$ is referenced on $A$, the total loss would increase if one would not account for the sensitivity of $B$.}.

Here lies another key point of differentiable connected geometries. If a functional and differentiable connection between the components is established, it can be differentiated by an auto differentiation framework. This ensures that gradients from child geometries are appropriately taken into consideration for the optimization of the parameters of the parent. The structure of the computational graph follows the structure of the connected geometry with one key difference. While the computational graph can follow a tree-like structure with multiple independent leaf-geometries, the geometries themselves need to follow a strict sequential ordering to avoid conflicts as depicted in \autoref{fig:atomic_geometries_conflict}. In the shown example, both $B$ and $C$ share no direct connection and can move and deform their free boundary independently from each other. This can lead to the depicted conflict region where it is unclear which of the geometries is finally realized. Imposing an ordering eg. $B$ becomes the parent of $C$, resolves the conflict without changing the underlying functional dependency. Both geometries deform independently but only on their respective free boundary. This sequential ordering can be interpreted as a priority-order on the realization of the geometry.

The ability to compute set operations with the geometry is required for any implementation of the connected geometries, especially for the distinction between restricted and free boundary. Deciding whether a point is in- or outside of its parent geometries is important as can be seen in \autoref{fig:atomic_geometries_conflict}. The process of building a dCG is similar to constructive solid geometry (CSG) \cite{foley:1996:Computer_Graphics:_Principles_and_Practice}, a technique used frequently in computer graphics and computer aided design (CAD). 



\subsection{Connected geometry gradients}

\begin{figure}[h]
  \centering
  \includegraphics[angle=0, trim = 0.5cm .5cm 0.5cm 2.2cm, clip, width = \columnwidth]{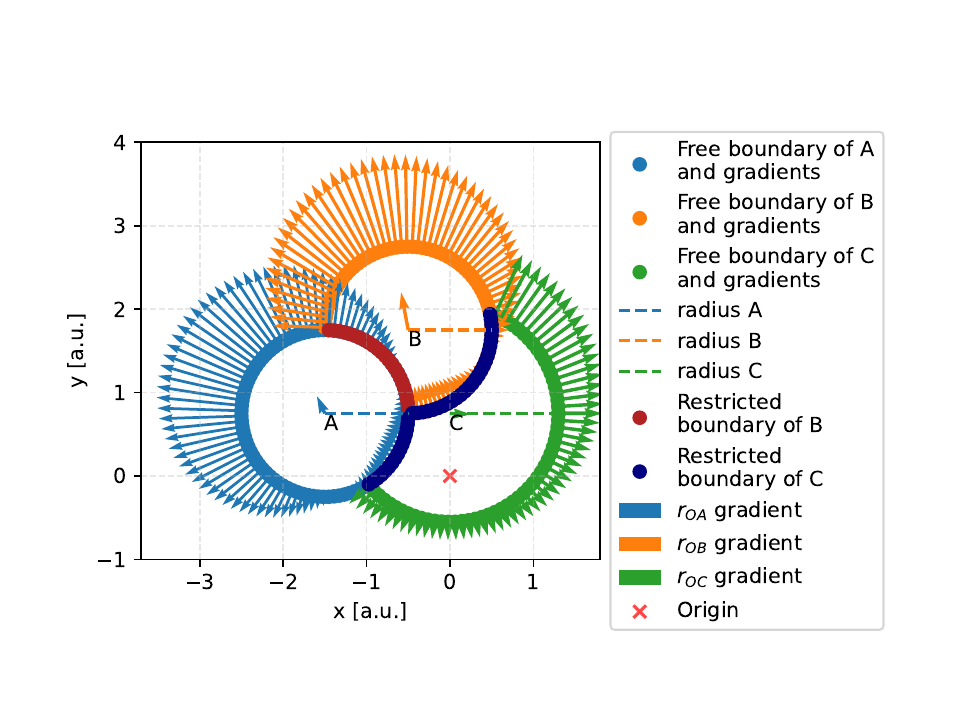}
  \caption{Implementation of the previously shown connected geometry with three circles. The loss function $\mathcal{L} = \sum_i \int \ds\, ||\p_\text{free} G_{i}||^2$ is evaluated on the geometry. The arrows indicate gradients, deforming the geometry in the opposite direction of the arrow decreases the loss. The gradient in the center indicates how the entire geometry should be moved to decrease the loss. Note that we only consider the size of the free boundary in this case. Different material properties are not taken into account. \label{fig:circle_gradients}}
\end{figure}

\begin{figure*}[ht]
    \centering    
    \begin{subfigure}[ht]{.25\textwidth}
        \centering
        \includegraphics[height=4.4cm, trim = 1.8cm 1.6cm 6.1cm 2.2cm, clip]{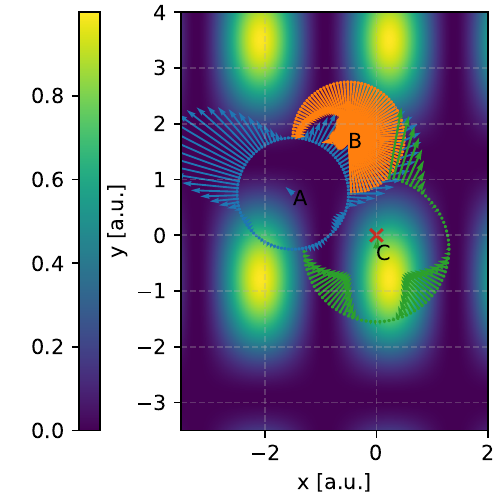}
        \subcaption{Initial circle geometry shown with the underlying function $f$. }
        \label{fig:circles_gradient_field_iter=0} 
    \end{subfigure}
    \hfill
    \begin{subfigure}[ht]{.35\textwidth}
        \centering
        \includegraphics[height=4.45cm, trim = 4.cm 1.6cm 1.2cm 2.2cm, clip]{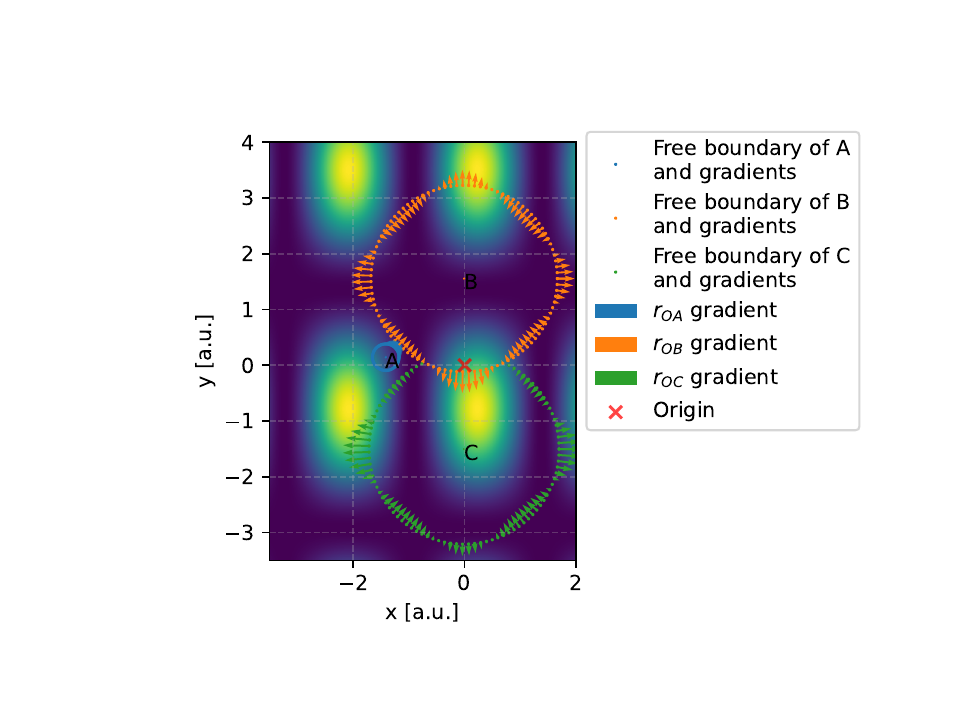}
        \caption{Geometry after 200 iterations. The geometry converged to a local minimum.}
        \label{fig:circles_gradient_field_iter=200}
    \end{subfigure}
    \hfill
    \begin{subfigure}[ht]{.33\textwidth}
        \centering
        \includegraphics[height=4.5cm, trim = .4cm 0.cm 0.cm 0cm, clip]{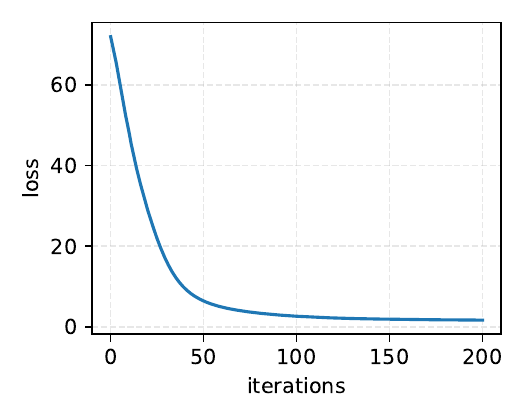}
        \caption{Loss function over the duration of the optimization. The loss shows good convergence behavior.}
        \label{fig:circles_gradient_field_loss}
    \end{subfigure}
    \caption{Three circle geometry optimization. The loss function $\mathcal{L} = \sum_i \int_{\p_\text{free}G_i} ds f(x,y) $ is evaluated for the loss with $f(x,y) = \bigl(\cos(x) + \sin(y)\bigr)^2$. The optimization is done via gradient descent with fixed learning rate.  \label{fig:example_optimization}}
    \vspace{.3cm}
\end{figure*}

Typically, the geometry in question is embedded into a computational domain that is governed by a set of partial differential system equations, ie. the physics governing the underlying dynamics. Often, one is interested in the sensitivity of the geometry wrt. a loss value. For example, decreasing the drag coefficient for the aforementioned airplane wing or finding a geometry for minimum reflection between two photonic waveguides. In shape calculus \cite{Delfour:2011:Shapes_and_Geometries}, the idea is to deform the boundary to gradually decrease the loss. The gradients are oriented parallel to the surface normal and their magnitude is proportional to the deformation. The gradients can be computed via the adjoint method \cite{Lebbe:2019:Contribution_in_topological_optimization_and_application_to_nanophotonics, Johnson:2021:Notes_on_Adjoint_Methods, Delfour:1991:Velocit_Method_and_Lagrangian_Formulation_for_the_Computation_of_the_Shape_Hessian}, which works by solving the system equations on the entire computational domain to obtain the s.c. \textit{forward} solution $u$ and repeating the computation again with the adjoint system equations to obtain the \textit{adjoint} or \textit{backward} solution $v$. Following \cite{Lebbe:2019:Contribution_in_topological_optimization_and_application_to_nanophotonics}, the optimal deformation $\delta \Omega$ on any domain $\Omega$ with boundary $\p\Omega$ is given by the variation of the loss functional $\mathcal{L}$
\begin{align}
    \delta \mathcal{L}_\Omega(\delta \Omega) = \int_{\p\Omega} \ds\, \delta \Omega \, \hat{n}\, V_{\Omega}(s)
\end{align}\label{eq:shape_variation} 
where $\hat{n}$ denotes the surface normals and the gradient field $V_\Omega = c(u\cdot v)$ is given by the forward and adjoint solutions of the system dynamics. The constant $c$ is domain dependent, in photonics for example, it depends on the permittivity and permeability of the domain materials \footnote{For the parallel electric field components and the respective domain permittivity: $c = \epsilon_1 - \epsilon_2$ and for the normal field components: $c = 1/\epsilon_1 - 1/\epsilon_2$. For the magnetic field components $c$ is given analogously by replacing the permittivity with the permeability.}.
Taking $\delta \Omega = - \hat{n}V_\Omega$ is guaranteed to decrease the loss functional for a small variation. 

Applying the shape gradients to a connected geometry is straightforward. For any geometry, the gradient field is only evaluated on the free boundary, the restricted boundary must not be evaluated and taken into account for the variation. For a connected geometry with a loss functional, we obtain gradients on the boundary of the geometry $\{ \delta\Omega(x) | x \in \p_\text{free}\Omega\}$. 

To demonstrate the optimization of connected geometries we present a simple example analogous to the shown arrangement of circles in \autoref{fig:circle_gradients}.
The three circle geometry boundaries  $\p G_i = \{\p A, \p B, \p C\}$ are instantiated via an autodifferentiation framework \footnote{We are using PyTorch as the autodifferentiation framework of choice \cite{Paszke:2017:Automatic_differentiation_in_PyTorch}.} and connected by their relative position vector. The functional connection is autodifferentiable and is thus taken care of by the autodifferentiation framework. In \autoref{fig:circle_gradients}, three circle geometries are instantiated and connected. Then, the loss $\mathcal{L} =\sum_i \int \ds\, ||\p_\text{free} G_{i}||^2$ is computed. The boundaries must be discretized for a numerical framework which can be done by either specifying a total number of boundary points or a stepsize. we choose a discretization of 100 points per boundary. By recording the gradients on the discretized points, we can directly show the aforementioned optimal boundary deformation $\delta\Omega$ on the geometry boundaries. By limiting the loss computation to the free boundary of each geometry we make sure to only account for the appropriate gradients wrt. the imposed ordering of the geometries. Finally, we would like to know how to change the geometry parameters ie. the position vector and radius of the three geometries in order to minimize the loss. The autodifferentiation framework backpropagates the gradients back to the original geometry parameters. The gradients of the position vector is shown in \autoref{fig:circle_gradients} in the center of each of the circles. Importantly, the gradient of $r_{OA}$ is informed about the change of the loss function by the child geometries $A$ and $B$ via backpropagation while the radius of all geometries is an independent parameter.

A key difference to conventional backpropagation is the behavior of the discretization methods. Typically for a discretization method such as  $f = \text{range}(x_\text{min}, x_\text{max}, n_\text{discretization}) = \{x_0 \dots x_n\}$ the default behavior of the gradient backpropagation is summing the gradients of the components: $\nabla f = \sum_i \nabla x_i $. In most cases, this is the correct behavior since the values $x_i$ are considered to be independent. Here, we need to be more careful since we must take into consideration that the created points are forming the boundary of a geometry. The total gradient is given by \autoref{eq:shape_variation} and we need to take the surface element $\ds$ into account. In the shown example, we modified the behavior of the discretization function such that it computes the Riemann sum during backpropagation $\nabla f = \sum_i \nabla x_i \Delta s$. Other, more precise numerical approximations of the integral are of course possible but it is important to keep in mind that many integration techniques such as the trapezoidal rule impose boundary conditions on the integral approximation. This can lead to artifacts of the gradients at the edges of the discretized array.

\subsection{Toy example optimization}\label{sec:example_opitmization}
A more realistic example would be to position the geometries inside a function such that the function value on the boundary is minimal. For example $f(x,y) = \bigl(\cos(x) + \sin(y)\bigr)^2$ with a loss function $\mathcal{L} = \sum_i \int_{\p_\text{free}G_i} \ds f(x,y) $. This example is closely related to the application of inverse design in a physical setting where a gradient field is known on the entire domain and the geometry is following the gradient field to optimize the loss\footnote{The difference in the presented example and a real application is that the gradient field changes with every iteration while we keep the example function $f$ constant for the sake of simplicity.}. We select SGD for the optimization and optimize circle position and radius jointly. The initial geometry is shown in \autoref{fig:circles_gradient_field_iter=0}. Then, 200 iterations of the optimization is performed. The geometry at the end of the optimization is shown in \autoref{fig:circles_gradient_field_iter=200}, with the corresponding loss shown in \autoref{fig:circles_gradient_field_loss}. At the end of the optimization, we see that the radius of both $B$ and $C$ increased and $A$ was reduced. The optimization found a local minimum of the loss function where $B$ sits between two maxima and $C$ encircles a maximum of $f$. More intermediate steps are shown in \autoref{appendix:intermediate_steps}. Complex geometry constraints can be implemented by either modifying the connection function of adding geometry regularization to the loss, for example a regularization term that penalizes too small or too large radii.

\subsection{Using dCGs for a photonic inverse design toy problem}\label{sec:application_to_photonic_inverse_design}

An interesting domain for geometric optimization is the field of photonic devices, where the development of \textmu LEDs and NanoLEDs is currently being advanced by researchers and industry \cite{Li:2023:Significant_Quantum_Efficiency_Enhancement_of_InGaN_Red_Micro-Light-Emitting_Diodes_with_a_Peak_External_Quantum_Efficiency_of_up_to_6, Vogl:2023:Role_of_pixel_design_and_emission_wavelength_on_the_light_extraction_of_nitride-based_micro-LEDs, Yan:2023:Enhanced_light_extraction_efficiency_of_GaN-based_green_micro-LED_modulating_by_a_thickness-tunable_SiO2_passivation_structure, Lex:2024:Impact_of_crystallographic_facet_of_InGaN_micro-LED_sidewalls_on_electro-optical_characteristics}. \textmu LEDs are being developed as a replacement for OLEDs due to their theoretical advantages \cite{Taki:2019:Visible_LEDs:More_than_efficient_light} over the latter. At its core, an epitaxially grown \textmu LED is a set of connected regions with different material properties, which influence the propagation of light within the device. However, growing \textmu LEDs epitaxially enforces certain geometric constraints on how individual layers can deform. Additionally, due to the sequential epitaxial growth process, modifying a layer deep inside the epitaxial stack influences the shape of the subsequent layers — a dependency which is difficult to take into account with traditional optimization approaches.

\begin{figure}[h]
  \centering
  \includegraphics[angle=0, trim = 0cm 0cm 0cm 0cm, clip, width = .99\columnwidth]{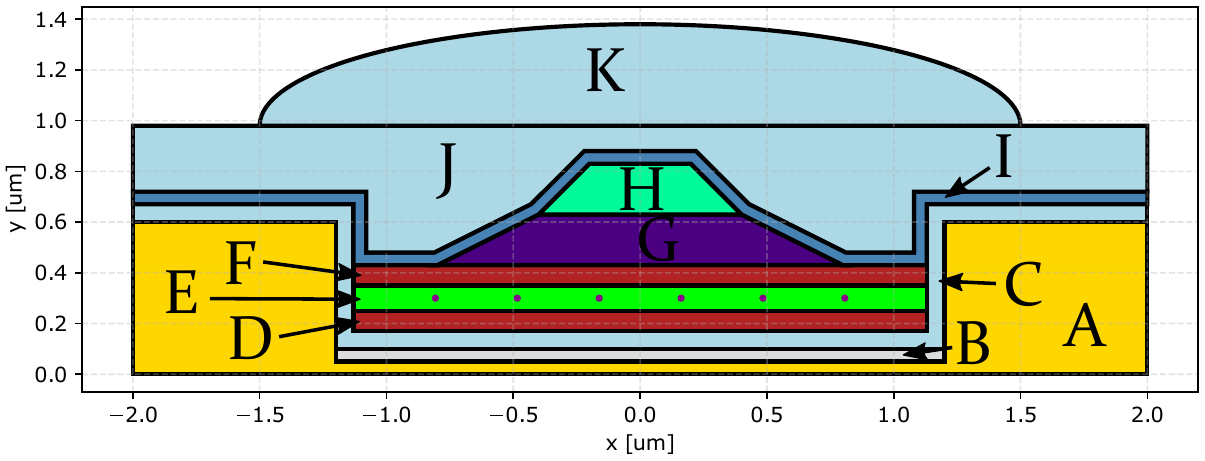}
  \caption{Cross-section of a toy model illustrating a micro-LED (\textmu LED) with parent-child ordering (indicated by letters, starting with A). Different layers, represented by distinct colors, exhibit varying light propagation properties due to their individual materials. The violet dots denote emitters within the \textmu LED’s active area. Layers are parameterized by thicknesses, angles, widths, and heights. Notably, the base geometry of the substrate significantly impacts subsequent layers, making it a crucial consideration during optimization.\label{fig:dCG_uLED_application}}
\end{figure}

Identifying free and restricted boundaries is straightforward in the model shown in \autoref{fig:dCG_uLED_application}. Due to the sequential nature of the epitaxial growth process, it is possible to define the free boundary as the boundary which is "open" in the upward direction. Any new layer receives the global free boundary (GFB) from its parent as input (see \autoref{appendix:connected_geometry_definitions}) and defines its own free boundary on top of the GFB, replacing the part of the GFB with the newly created free boundary. This approach mimics the growth process of the device and naturally results in the correct assignment for free and restricted boundaries.

Since a new child geometry receives the GFB as input, it can apply a function to the GFB depending on the type of geometrical layer which should be added — for example, in \autoref{fig:dCG_uLED_application} a flat layer (B, D, E, F, J), a displacement layer (C, I), a trapezoid layer (G, H), or a freeform lens layer (K). By using an autodifferentiable function to create the new geometry, we naturally obtain a computational graph which can backpropagate gradients back through the model.

From here, it remains to compute the physical gradients of the system. For a photonic system, this involves solving Maxwell's equations in the device given the sources inside the active region. A common target is finding a geometry which maximizes the radiated power into a particular direction, given a fixed optical input power — that is, maximizing the light extraction efficiency (LEE) into a particular solid angle $\Gamma$. Hence, the loss function is given by
\begin{align}
    \mathcal{L} = -\int_{\pm \Gamma}\varrho_\text{LEE}(\theta)\, \d\theta
\end{align}
where $\varrho_\text{LEE}(\theta)$ is the so called LEE density, see \autoref{appendix:uLED_optimization_with_dCG}.

\begin{figure*}[ht]
    \centering    
    \includegraphics[width=0.99\textwidth, trim = 0cm 0.cm 0.cm 0.cm, clip]{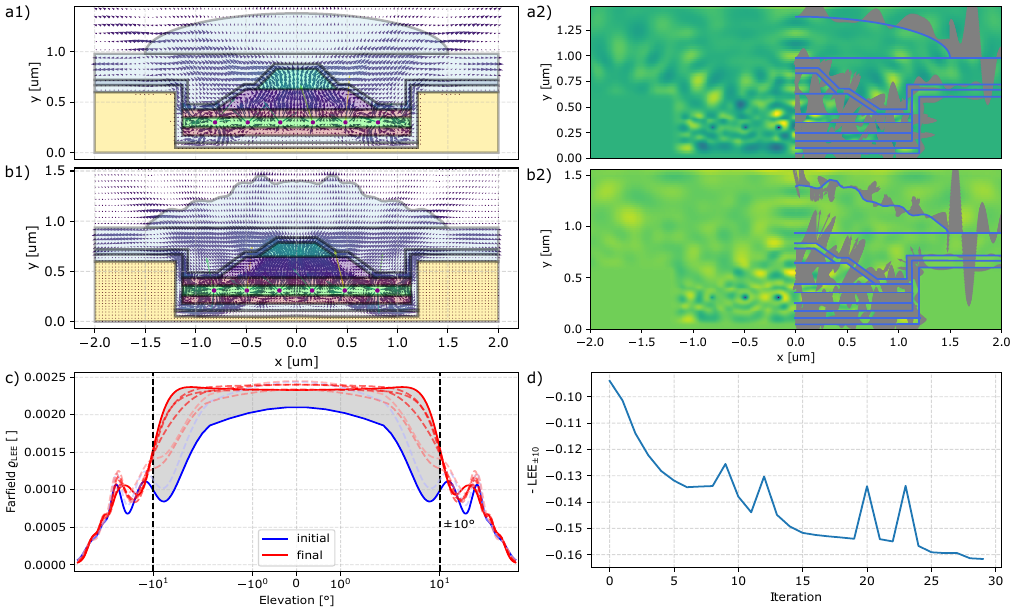}
    \caption{Overview of the optimization of the toy \textmu LED structure shown in \autoref{fig:dCG_uLED_application}. In a1), we show the power flow, i.e., the time-averaged Poynting vector within the \textmu LED geometry, emitted from dipole sources embedded in the structure. The goal is to reshape the internal structure of the \textmu LED such that more power is radiated into the $\pm$10° far-field solid angle. Due to the wave-optical nature of the underlying physics, optimizing this system is challenging even for experts. By solving the adjoint equation \cite{Luce:2024:Merging_automatic_differentiation_and_the_adjoint_method_for_photonic_inverse_design}, we obtain the sensitivity field, which is subsequently evaluated on the boundary of the geometry, shown in a2). The value of the gradient field, together with the boundary normal, results in the shape gradients, which are shown in gray. Note that the shape gradients are evaluated with a very fine discretization and thus appear as a continuous area. A zoomed version of the shape gradients is shown in \autoref{appendix:uLED_optimization_with_dCG}. By backpropagating the gradients through the dCG model, we obtain the parameter gradients, which are used to perform gradient descent on the geometry. After 30 steps, the loss shown in d) approaches a value of $\mathcal{L}=-0.16$, which indicates an improvement of $72\%$ with respect to the initial design. The far-field LEE densities during the optimization are shown in c), illustrating the improvement in the emission in the $\pm$10° direction. The highlighted area represents the improved density during the optimization and is proportional to the improvement of the loss. The improvement can also be seen in the power flow shown in b1), where a larger share of the Poynting vectors are now oriented upwards. Note that the shape gradients are still present on the structure in b2); this indicates that only a local minimum for the parameters was found, since the shape gradients are now oriented such that they will cancel each other when backpropagating to the parameters. Relaxing the requirements of strictly satisfying the parameters might result in further optimization and subsequently an even better result. \label{fig:dCG_uLED_optimization}}
\end{figure*}

To obtain the solution to Maxwell's equations, one can utilize different numerical solvers, such as FDTD \cite{lumerical}, FDFD \cite{Vuckovic:2019:Nanophotonic_Inverse_Design_with_SPINS:_Software_Architecture_and_Practical_Considerations}, or RCWA \cite{Vial:2023:Nannos}. For this application we used Lumerical as FDTD solver. To differentiate the FDTD simulation, we used the Adjoint Method \cite{Johnson:2021:Notes_on_Adjoint_Methods, Hansen:2015:Accurate_adjoint_design_sensitivities_for_nano_metal_optics, Lebbe:2019:Contribution_in_topological_optimization_and_application_to_nanophotonics} and integrated the FDTD solver directly into the computational graph \cite{Luce:2024:Merging_automatic_differentiation_and_the_adjoint_method_for_photonic_inverse_design} of the autodifferentiation framework we employed \cite{Paszke:2017:Automatic_differentiation_in_PyTorch} for the geometry creation. 
This enables gradients to seamlessly flow through the entire chain of steps from the loss function down to the parameterization. The output of the adjoint method is the so called sensitivity, or velocity field. This field represents the optimal change\footnote{To first order.} of the material distribution. Evaluating the field on the free boundary of the layers and taking into account the shape gradient \cite{Delfour:2011:Shapes_and_Geometries, Lebbe:2019:Contribution_in_topological_optimization_and_application_to_nanophotonics, Luce:2024:Merging_automatic_differentiation_and_the_adjoint_method_for_photonic_inverse_design} we obtain boundary gradient vectors which indicate how a particular point on the boundary needs to be displaced in order for the loss function to decrease. The target farfield as well as the sensitivity field and the resulting shape gradients can be seen in \autoref{fig:dCG_uLED_optimization}. Since we are not interested in how to displace a single point of the boundary but rather the entire boundary at once, it remains to integrate the boundary gradients to figure out which net displacement is beneficial. Fortunately, since the boundary discretization is generated by an autodifferentiable function this integration is automatically handled by the autodiff framework as well as the dependency of the discretization on the geometry parametrization. The parameter gradients can now be backpropagated back through the model. Thus, the parameter gradients in the first parent layers takes into account all changes to the total geometry which would result by changing it. The results of an examplary optimization of the given structure are shown in \autoref{fig:dCG_uLED_optimization} with the initial and final LEE densities shown in \autoref{fig:dCG_uLED_optimization} c). Further details of the optimization can be found in \autoref{appendix:uLED_optimization_with_dCG}.



\subsection{Comments on differentiability violations}
\begin{figure}[ht]
  \centering
  \includegraphics[angle=0, trim = 0.cm 0.cm 0.cm 0.cm, clip, width = .5\columnwidth]{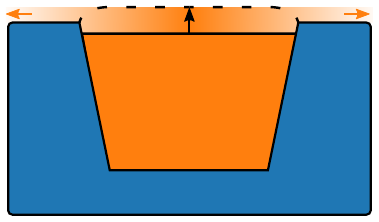}
  \caption{Potential non-differentiable deformation - if the planar boundary of the orange interior is displaced upwards, following the black arrow, the boundary must deform in order to be differentiable at the upper edge of the blue geometry. A potential valid deformation is shown by the dashed line where the upper boundary follows the contour of the blue shape. \label{fig:continuous_deformation}}
\end{figure}

A key requirement for the definition of a differentiable geometry is that all geometric changes that occur must be smooth and differentiable. Any infinitesimal distortion of the parameters must lead to an infinitesimal change of the boundary. Otherwise, the geometry is not differentiable. In \autoref{fig:continuous_deformation}, a situation is shown in which a potential non-continuous deformation could be performed. If the upper boundary of the interior geometry is shifted upwards, a non-differentiable deformation could be performed if the boundary is kept straight. A small increase in height would lead to a large change of the geometry. Such geometry definitions can be useful nonetheless since the geometry is differentiable for downwards deformations of the boundary. In this case, it is paramount to make sure that the interior geometry never crosses the upper boundary of the surrounding geometry. The alternative is of course to utilize geometries which can only deform continuously, such as TDFs \cite{Sigmund:2013:Topology_optimization_approaches}.

In addition to smooth deformations it is important to compute a loss which can be appropriately handled by the geometry definition. For instance, for a situation similar to \autoref{fig:atomic_geometries} if the loss function computes an integral of a function $f$ on the boundary of $A \cup B$, the gradient from the two connected geometries with internal free boundary will not be correct since the union of the geometry is never computed explicitly. Increasing the radius of B would therefore change the boundary of $A \cup B$ in a way that influences the loss without taking into account that parts of the surface from $A$ will be covered. Gradients will be computed on the entire free surface of the domain, including the interior free boundaries. 

\begin{figure}[ht]
  \centering
  \includegraphics[angle=0, trim = 0.cm 0.cm 0.cm 0.cm, clip, width = .8\columnwidth]{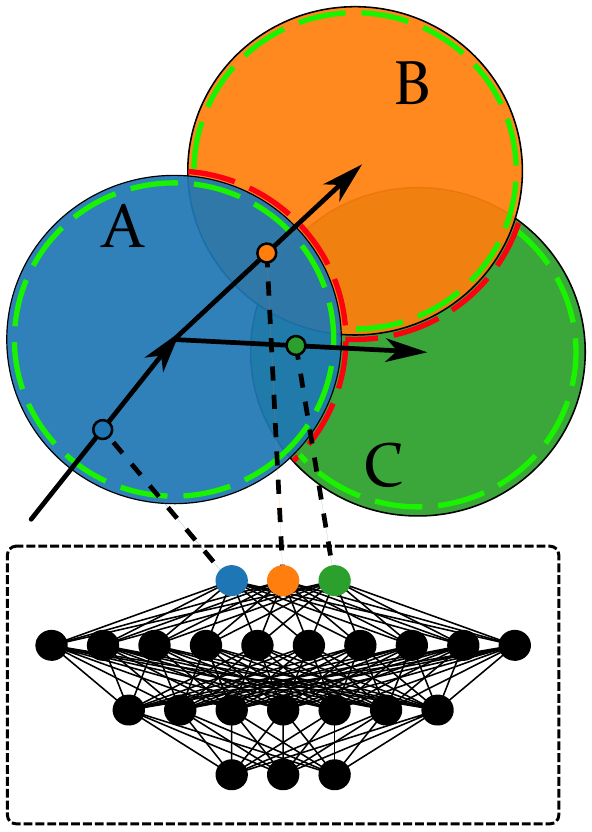}
  \caption{Connecting dCGs with neural networks is straightforward. The boundary discretization or the geometry parameterization can be directly interpreted as the output or input to a neural network. The gradients can therefore backpropagate from the loss of the physical system directly to the neural network weights. \label{fig:neural_networks_and_connected_geometries} \vspace{-.7cm}}
\end{figure}
\subsection{Guarantees on the constraints of the parameter gradients}
Constraints and following the correct and desired geometry deformation is an important requirement for any optimization tool. In parameterized geometric optimization, which should generate manufacturable geometries, it is particularly important to obey geometric constraints. However, gradients are generally agnostic to hard constraints and often, they can only be weakly enforced via regularization by Lagrange multipliers. 

Here, dCG offers a significant advantage compared to other geometric optimization problems because the degree of strict following of constraints and flexibility is given to the user. Given an explicit definition of a geometry, the update is directly performed on the explicit parameterization of the geometry which inherently obeys the constraints posed in the geometry definition. For example, updating the radius of a circle will always result in a new circle being created, albeit with a different size. A more detailed explanation is given in \autoref{appendix:guarantees_on_constraints}.


To enhance flexibility during the optimization process, we can introduce additional optional parameters $\varepsilon$ to augment the geometry parameters. This parameter allow a displacement additionally to the one permitted by the constraints as shown in \autoref{appendix:constraint_violation}. Importantly, $\varepsilon$ is also an optimizable parameter. By doing so, we can account for the sensitivity associated with violating the constraints.

To maintain fidelity to the constraints, we gradually decay $\varepsilon$ to zero as the optimization progresses. Allowing the geometry to violate the constraints temporarily can be advantageous during optimization. It helps the optimization algorithm avoid getting stuck in local minima, even if it comes at the cost of potentially increasing the overall loss to meet the constraints.


\label{sec:NN_controller}
\subsection{Combining differentiable connected geometries with machine learning}

\begin{figure*}[ht]
    \centering
    \begin{subfigure}[ht]{.25\textwidth}
        \centering
        \includegraphics[height=4.5cm, trim = 1.8cm 1.6cm 6.1cm 2.2cm, clip]{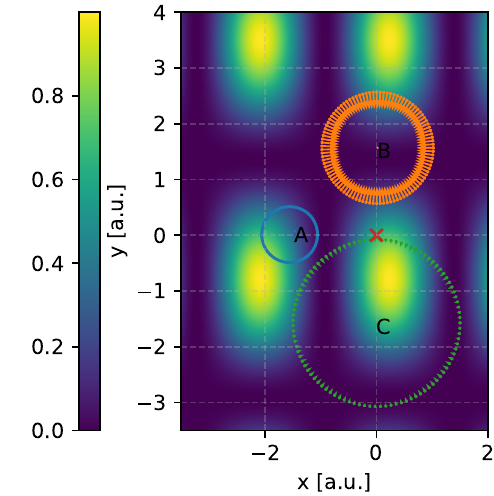}
        \caption{Geometry after 3000 iterations, optimized via SGD. }
        \label{fig:sgd_gradient_field_iter3000} 
    \end{subfigure}
    \hfill
    \begin{subfigure}[ht]{.35\textwidth}
        \centering
        \includegraphics[height=4.55cm, trim = 4.cm 1.6cm 1.2cm 2.2cm, clip]{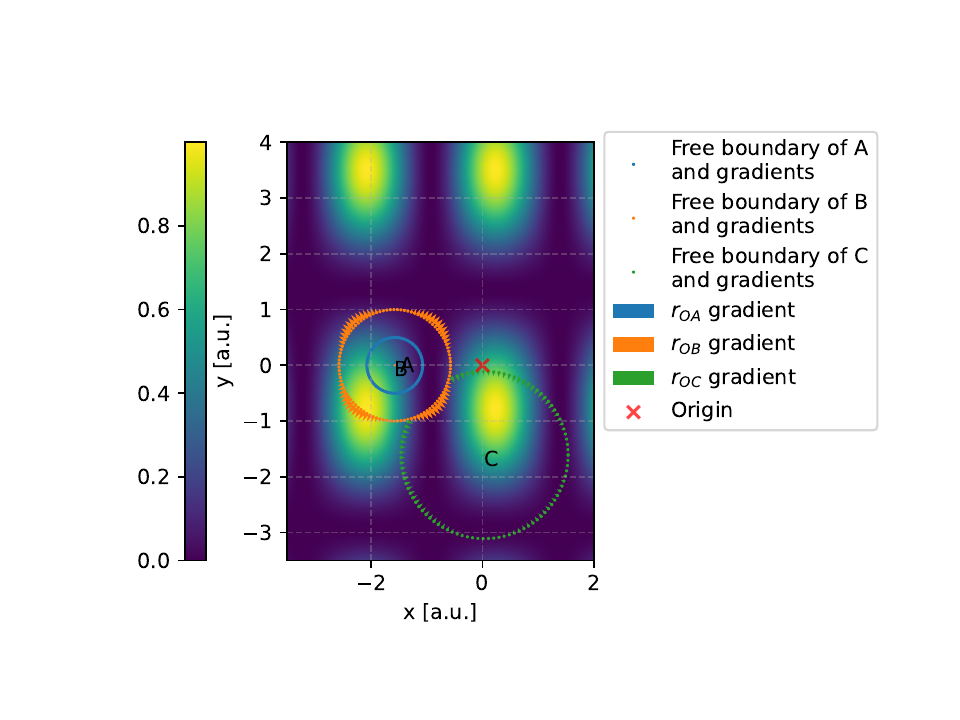}
        \caption{Geometry after 3000 iterations, optimized via GloNet. }
        \label{fig:glonet_gradient_field_iter3000}
    \end{subfigure}
    \hfill
    \begin{subfigure}[ht]{.33\textwidth}
        \centering
        \includegraphics[height=4.8cm, trim = .4cm 0.cm 0.cm 0cm, clip]{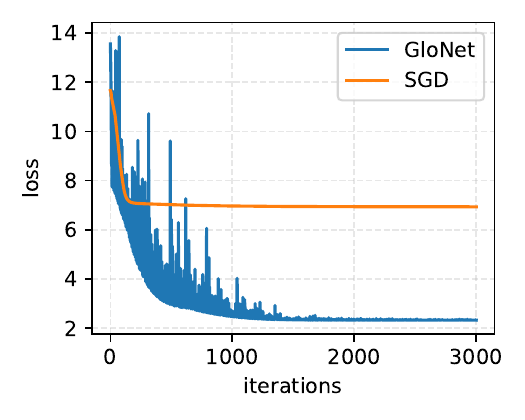}
        \caption{Optimization loss comparison between SGD and a GloNet. }
        \label{fig:circles_loss_glonet_vs_SGD}
    \end{subfigure}
    \caption{Comparison of the optimization of the circle geometries with SGD and with a GloNet. For SGD, the starting geometry is the same as in \autoref{fig:circles_gradient_field_iter=0}. For the GloNet, the starting geometry depends on the sampled latent space vector which is random. Here, the optimization parameters were the position vectors of the circles only and the radius was kept constant. Additionally, the loss was modified by a regularization term of the length of the position vector to avoid diverging geometries. \label{fig:sgd_vs_glonet}}
\end{figure*}

For machine learning applications, differentiable connected geometries offer a particularly useful advantage: coupling the geometry to a neural network is straightforward. The parameterization of the connected geometry can be directly passed to a neural network, or the output of the neural network interpreted as the parameterization of the geometry. Since the geometry itself is differentiable, the network becomes part of the evaluation process. This simplifies the process of combining neural networks with topology optimization techniques compared with previous approaches where the optimization domain was connected pixelwise to a neural network \cite{Nie:2021:TopologyGAN:_Topology_Optimization_Using_Generative_Adversarial_Networks_Based_on_Physical_Fields_Over_the_Initial_Domain}. Having a differentiable geometry definition might be especially useful for optimal control or policy gradient reinforcement learning in complex systems. The dCG approach offers another element in a set of differentiable components for complex systems \cite{Amos:2018:Differentiable_MPC_for_End-to-end_Planning_and_Control, de_Avilla:2018:End-to-End_Differentiable_Physics_for_Learning_and_Control}.

For demonstration purposes, we implemented a very simple version of a so-called GloNet\footnote{In the shown case, we used a 4 + 1 output layer fully connected neural network with ReLU activation and one dropout layer.} (global optimization network) \cite{Jiang:2019:Global_Optimization_of_Dielectric_Metasurfaces_Using_a_Physics-Driven_Neural_Network}. The model receives a random latent space vector as input. The network output is interpreted as the geometry parameterization from which the geometry is instantiated, evaluated, and finally a loss is computed. Then, the gradients are backpropagated through the geometry, and the GloNet weights can be updated. By sampling the input from a normal distribution, the GloNet samples through many different input parameterizations and learns the loss landscape of the underlying system, hence the name global optimization network. A converged GloNet maps all points from the latent space distribution onto the best set of geometry parameters. A comparison between an optimization with plain gradient descent is shown in \autoref{fig:circles_loss_glonet_vs_SGD} with the loss function $\mathcal{L} = \sum_i \int_{\p_\text{free}G_i} ds f(x,y)  + \frac{1}{N}\sum_i^N \text{ReLU}(||\Vec{r_i}||-5)$. The second term acts as a regularization such that the distance between the geometries $\Vec{r_i}$ is penalized for large separation.

We observe the expected result: gradient descent quickly reduces the loss until it gets stuck in a local minimum, while the GloNet takes more iterations to converge but achieves a better optimum than SGD. The minima after 3000 iterations are shown in \autoref{fig:sgd_gradient_field_iter3000} and \autoref{fig:glonet_gradient_field_iter3000}.


\section{Discussion and Limitations}
\label{sec:4_limitations}
Considering the limitations of the approach, it is important to mention the computational complexity of Boolean operations, which are necessary to distinguish free and restricted parts of the boundary. Particularly, for very large 3D problems, determining the boundaries can become computationally infeasible. At its core, the complexity scales with the complexity of the so-called Point-in-Polygon problem \cite{Hormann:2001:The_point_in_polygon_problem_for_arbitrary_polygons, Greiner:1998:Efficient_clipping_of_arbitrary_polygons} which is $O(n)$ with $n$ the number of polygon edge points. Since it is necessary to check all points that make up the geometries to determine whether they are part of the free or restricted boundary, the scaled complexity becomes $O(n^2)$ at worst for arbitrary polygons. However, many specializations \cite{Trettner:2021:Sampling_from_Quadric-Based_CSG_Surfaces, Trettner:2022:EMBER:_exact_mesh_booleans_via_efficient_&_robust_local_arrangements}, exist which can significantly speed up determining whether a point is free or restricted\footnote{E.g., for the geometry in \autoref{fig:circle_gradients} , it suffices to check the distance of any given point to the center of the parent geometries, thus reducing the complexity to $O(p\,n)$ with $p$ the number of parent geometries.}. For the problem of computing intersections for boundary representations or constructive solid geometry, many different approaches already exist and are widely used in commercial and open-source tools such as CAD programs or computational geometry solvers \cite{cgal:2024:Boolean_Operations_on_Nef_Polyhedra, Trettner:2022:EMBER:_exact_mesh_booleans_via_efficient_&_robust_local_arrangements, Bernstein:2016:Cork_Boolean_Library}. 

For many scientific and engineering optimization problems, however, the cost of computing the free and restricted boundaries should be dwarfed by the computational cost of finding the correct gradients\footnote{The time to construct the dCG model for the photonic optimization problem described in \autoref{sec:application_to_photonic_inverse_design} is negligible compared with the computational cost of solving the adjoint equations to compute the gradients.}

Finally, the optimization is very sensitive to hyperparameter tuning. Setting the learning rate incorrectly will result in parameters reaching their bounds quickly, performing non-differentiable geometry changes if the geometry creation framework cannot appropriately handle the change, or not changing the boundary at all, which results in a constant geometry. Additionally, the orders of magnitude of the individual parameters can vastly differ. In the toy \textmu LED optimization example, the angles in the geometry are on the order of $\sim 1$ rad, while the thicknesses are on the order of $[10^{-6} - 10^{-9}]$ meters. This entails tedious fine-tuning of learning rates and momentum parameters to achieve a consistent deformation of the geometry and subsequently better convergence to a minimum. Improving the convergence behavior and rescaling the parameters and gradients to achieve similar orders of magnitude should be the subject of future work.

\section{Conclusion}
\label{sec:4_conclusion}
This paper introduces differentiable connected geometries (dCGs) and shows how they can be used to integrate shape optimization into a computational framework that leverages automatic differentiation for optimization. We demonstrate how geometries can be systematically ordered and connected, ensuring that parent and child relationships are clearly defined to mitigate conflicts and maintain a hierarchy that respects geometric constraints.

Key to this framework is the distinction between free and restricted boundaries, which allows for the computation of gradients necessary for optimization tasks, ensuring that only the appropriate parts of the geometry contribute to the optimization process. The paper further demonstrates the practical application of this approach through examples, showing the potential of dCGs to refine and optimize geometric configurations in a computationally efficient manner. Importantly, the approach supports the integration of machine learning models, particularly neural networks, which can be directly connected to the geometry, allowing gradients to be backpropagated through. The compatibility of dCGs with existing machine learning frameworks was exemplified through the implementation of a Global Optimization Network (GloNet).

By providing a systematic way to define and optimize connected geometries, the framework lays the groundwork for future research and applications that bridge the gap between geometric design and machine learning. We hope that our contribution will be useful to other researchers and enable new and better optimization approaches in the field of shape and topology optimization.

\subsection*{Acknowledgments}
We thank the supporters of this work, particularly Heribert Wankerl (former ams OSRAM) and Maike Stern (OTH Regensburg) for constructive feedback of this manuscript. Furthermore, we thank Rasoul Alaee (ams OSRAM), Fabian Knorr (ams OSRAM), Johannes Oberpriller (ams OSRAM) and Philipp Schwarz (ams OSRAM \& University Regensburg) for fruitful and helpful discussions. Finally, we thank Harald Laux (ams OSRAM) for his organizational support.

\subsection*{Publication Funding Acknowledgments}
Funding from the German Federal Ministry of Economic Affairs and Climate Action (BMWK) and the Bavarian State Ministry of Economic Affairs and Media, Energy and Technology within the IPCEI-ME/CT “OptoSure (GA: 16IPCEI221) is gratefully acknowledged.

\section*{Impact Statement}
We presents work with the intention to advance the field of geometric optimization towards better designs and integration with deep learning. To the best of our knowledge, there are no social or ethical risks associated with this article.

\bibliographystyle{unsrt}  
\bibliography{0_references}  

\section{Appendix}
\label{sec:5_appendix}

\begin{figure*}[ht]
    \begin{subfigure}[ht]{.28\textwidth}
        \centering
        \includegraphics[height=5.4cm, trim = 3.2cm 1.6cm 6.4cm 2.2cm, clip]{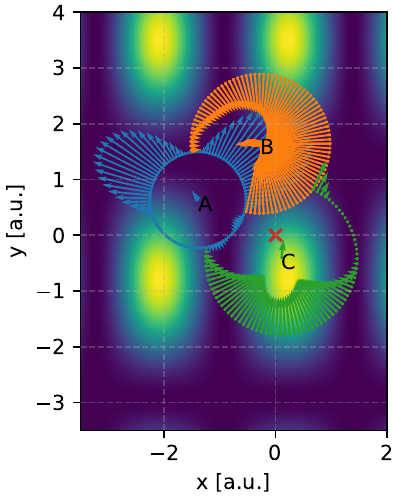}
        \caption{Geometry after 10 iterations.}
    \end{subfigure}
    \hfill
    \begin{subfigure}[ht]{.28\textwidth}
        \centering
        \includegraphics[height=5.4cm, trim = 4.cm 1.6cm 6.4cm 2.2cm, clip]{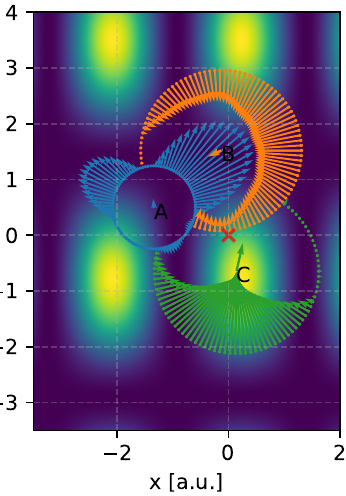}
        \caption{Geometry after 20 iterations.}
    \end{subfigure}
    \hfill
    \begin{subfigure}[ht]{.40\textwidth}
        \centering
        \includegraphics[height=5.45cm, trim = 4.cm 1.6cm 0.cm 2.2cm, clip]{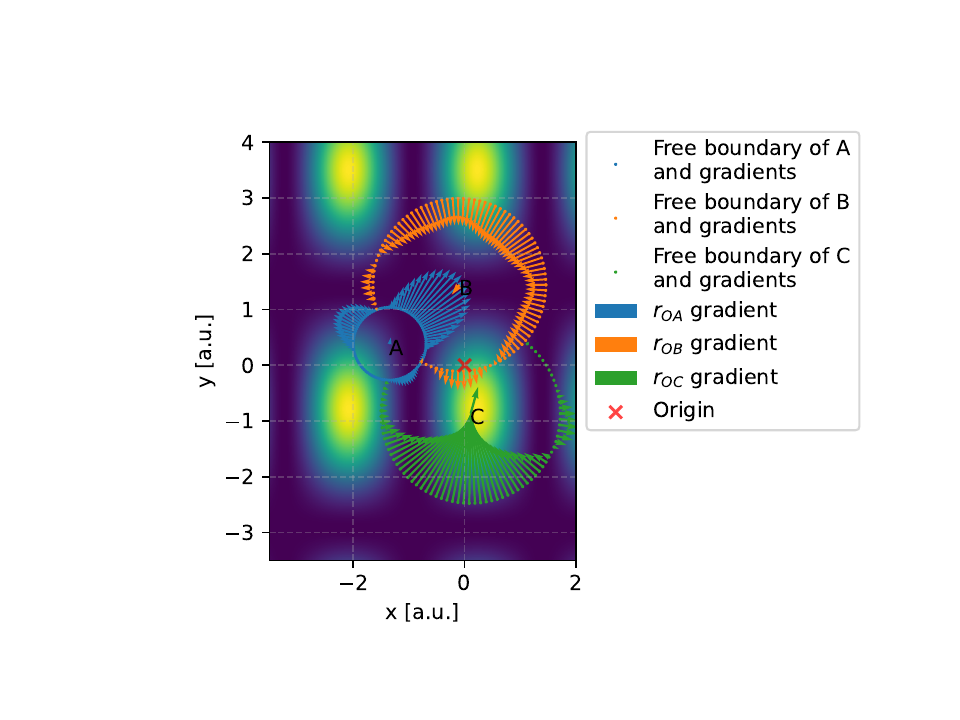}
        \caption{Geometry after 30 iterations. }
    \end{subfigure}
\end{figure*}

\begin{figure*}[ht]
    \centering
    \begin{subfigure}[ht]{.28\textwidth}
        \centering
        \includegraphics[height=5.4cm, trim = 3.2cm 1.6cm 6.4cm 2.2cm, clip]{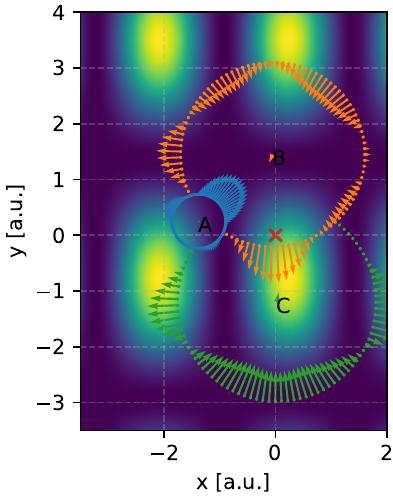}
        \caption{Geometry after 60 iterations.}
    \end{subfigure}
    \hfill
    \begin{subfigure}[ht]{.28\textwidth}
        \centering
        \includegraphics[height=5.4cm, trim = 4.cm 1.6cm 6.4cm 2.2cm, clip]{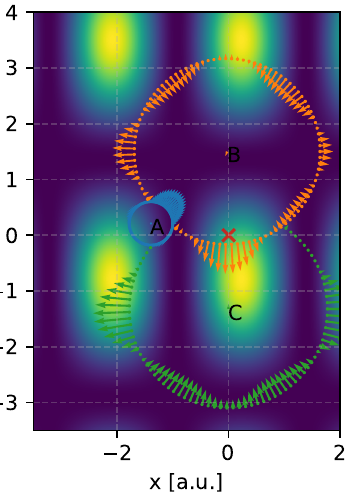}
        \caption{Geometry after 90 iterations.}
    \end{subfigure}
    \hfill
    \begin{subfigure}[ht]{.40\textwidth}
        \centering
        \includegraphics[height=5.45cm, trim = 4.cm 1.6cm 0.cm 2.2cm, clip]{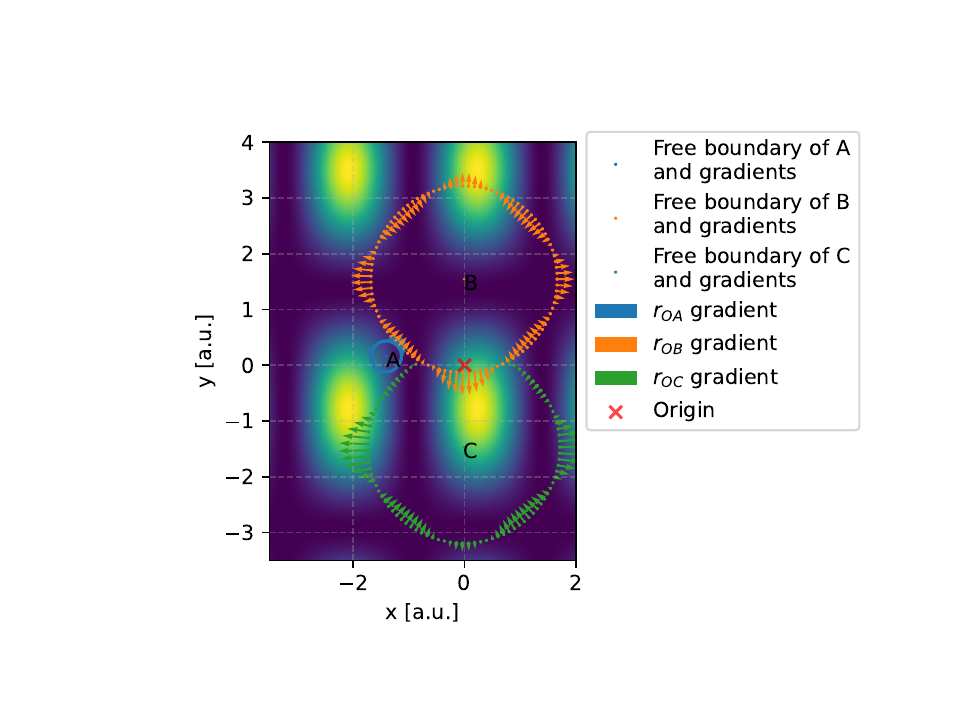}
        \caption{Geometry after 160 iterations. }
    \end{subfigure}
    \caption{Intermediate steps of the circle geometry optimization. \label{fig:circle_intermediate_steps}}
\end{figure*}

\subsection{Connected geometry definitions}
\label{appendix:connected_geometry_definitions}
Assigning the correct free and restricted boundaries lies at the core of dCG approach. Therefore, in the following, we provide the mathematical definitions for the different components of dCGs. For simple geometries $A_i \subset \mathbb{R}^n$ the boundary of the geometry is denoted by $\partial A_i$. In the paper, we assumed that $n=2$ due to simplicity.

Def: a free boundary, denoted by $\p_f$:
\begin{align}
    \bigl\{\p_f A_i \subseteq \partial A_i |\, & A_i(p'),\, p' = p_0 + \varepsilon \\&\Rightarrow \p A_i(p') - \p A_i(p) \neq 0\bigr\}. 
\end{align}
The free boundary indicates a completely free boundary where changes to any geometric parameters $p$ would result in a change of the subset of the boundary. Thus, the gradient needs to be considered. For a geometry without any parent $\p_f A_i = \partial A_i$, applies.

Def: a restricted boundary, denoted by $\p_r$:
\begin{align}
    \bigl\{\p_r A_i \subseteq \partial A_i |\, & A_i(p'),\, p' = p_0 + \varepsilon \\&\Rightarrow \p A_i(p') - \p A_i(p) = 0\bigr\}. 
\end{align}
denotes a part of the boundary which is independent from change to geometric parameters. 

The free boundary is the complement of the restricted boundary: 
\begin{align}
    \p_f A_i = \partial A_i / \partial_r A_i,
\end{align}
thus any part of the boundary $\p A_i$ is either free or restricted. 

Typically, a restricted boundary emerges when two geometries intersect. Consider two geometries $A_i$ and $A_j$ where $A_i$ is the parent of $A_j$ and $A_i$ has no parent. 
The restricted boundary of $A_j$ is given by 
\begin{align}
    \p_r A_j = \p A_j \cap A_i
\end{align}
If $\p_r A_j \neq \emptyset$, then we call the two geometries \textit{physically connected}.
Otherwise, if $\p_r A_j = \emptyset$, the geometries are \textit{not physically connected}. Thus, two geometries maintain their parent-child relationship even if they are separated such that they don't touch and subsequently have only free boundaries. 

In the case of more than two geometries, all parents of the parent must be taken into account to compute the restricted boundary:
\begin{align}
     \p_r A_j = \p A_j \bigcap_{i_p = 0}^N A_i
\end{align}
Note that any child geometry, however, can only have a single direct parent due to the sequential graph required to avoid conflict regions, see \autoref{fig:atomic_geometries_conflict}.
It is often useful to define a \textit{composed geometry} as a subset of the entire structure that has a specific purpose. For example, it can be useful to define a composed geometry for a set of stacked rectangles which have a total maximum height.
Def: a composed geometry is given by the union of a sequence of ordered and physically connected geometries: 
\begin{align}
    C = \bigcup_{i=0}^N A_i
\end{align}
with $A_{i+1}$ a child of $A_i$ for all $i$. 

A consequence of the way how the restricted boundaries are computed is that they will always be internal boundaries. Therefore, it is useful to compute the \textit{global geometry} (GG)

Def: the global geometry (GG): 
\begin{align}
    \text{GG} = \left( \bigcup_{i=0}^N C_i \right) \cup \left( \bigcup_{i=0}^M A_i \right)
\end{align}
with all composed and individual geometries $C_i$ and $A_i$. As the name suggests, the GG denotes all geometries in the domain. It is useful to keep track of the GG during computation. Due to the fact that only internal boundaries can be restricted, it is easy to see that the entire boundary of GG must be free which simplifies computing the free boundary when adding a new child geometry. The free boundary of the GG is therefore called 
Def: global free boundary (GFB)
\begin{align}
    \text{GFB} = \p_f \text{GG}
\end{align}

No matter to which geometry the child $A_j$ is connected, its restricted boundary is given by 
\begin{align}
    \p_r A_j = C_i \cap \text{GFB}.
\end{align}
and the free boundary follows directly by $\p_f \A_j = \p A_k / \p_r A_j$.

\subsection{Intermediate optimization steps}
\label{appendix:intermediate_steps}
Additional intermediate steps from the example optimization shown in \autoref{sec:example_opitmization} are depicted in \autoref{fig:circle_intermediate_steps}.


\subsection{\textmu LED optimization with dCG}\label{appendix:uLED_optimization_with_dCG}
In \autoref{sec:application_to_photonic_inverse_design}, we demonstrated the application to a photonic design problem optimization. Since the optimization involves multiple intermediate steps and additional results, they are presented here for completeness. 
The device geometry resembles the internal structure of a toy model \textmu LED.
\subsubsection{FDTD model}
For the sake of simplicity and optimization speed, we assume a 2D problem which implies a continuous extension of the 2D geometry in the perpendicular direction of the 2D cross section. For the individual layers / geometries of the FDTD model \cite{lumerical} we assumed the materials given in \autoref{tab:materials_and_parameters}. Geometry E (GaN) is assumed to be the light emitting layer, ie. the so called active area. For a \textmu LED, we must use incoherent and distributed dipole emission form all over the active area. Therefore, the emitter positions are distributed within the green GaN geometry and indicated by the violet dots \autoref{fig:dCG_uLED_application}. 
We assume an emission spectrum of the toy model \textmu LED with a gaussian weighting and peak wavelength at 625nm and FWHM of 20nm. 
\begin{table}[]
    \centering
    \begin{tabular}{c|c|c|c}
    \makecell{\#} & Material & color & \makecell{optimzable\\ parameter} \\
    \hline\hline
    A & gold (Au) & \makecell{yellow/\\golden} & \makecell{edge point \\ coordinates} \\
    B & \makecell{indium-tin-oxide\\ (ITO)} & white & \makecell{thickness}\\  
    C & \makecell{silicon dioxide\\ (SiO2)} & lightblue & \makecell{thickness}\\
    D & \makecell{aluminium-indium\\-gallium-phosphid\\ (InGaAlP)} & red & \makecell{thickness}\\
    E & \makecell{gallium nitrid\\ (GaN)} & lime & \makecell{thickness}\\
    F & \makecell{aluminium-indium\\-gallium-phosphid\\ (InGaAlP)} & red & \makecell{thickness} \\
    G & \makecell{niobium pentoxide\\ (Nb2O5)} & indigo & \makecell{prism height,\\lower \& upper \\width} \\
    H & \makecell{Silicon nitride\\ (Si3N4)} & lightgreen & \makecell{prism height,\\lower \& upper \\width}\\
    I & \makecell{aluminium oxide\\ (Al2O3))} & darkblue & thickness\\
    J & \makecell{silicon dioxide\\ (SiO2)} & lightblue & thickness\\
    K & \makecell{silicon dioxide\\ (SiO2)} & lightblue & \makecell{curve points \\ coordinates}\\
    \end{tabular}
    \vspace{.3cm}
    \caption{Material selection for the individual layers / geometries in the sandbox \textmu LED optimization problem and the parameters available for the optimization. Note that thickness refers to the thickness at y = 0 for the flat geometries (B, D, E, F, H) whereas for the contour following geometries it means the thickness perpendicular to the underlying contour (C, G). }
    \label{tab:materials_and_parameters}
\end{table}

\begin{figure*}[ht]
    \centering
    \begin{subfigure}[ht]{.525\textwidth}
        \centering
        \includegraphics[height = 9.5cm, trim = 2.8cm 0.cm 2.8cm 0.cm, clip]{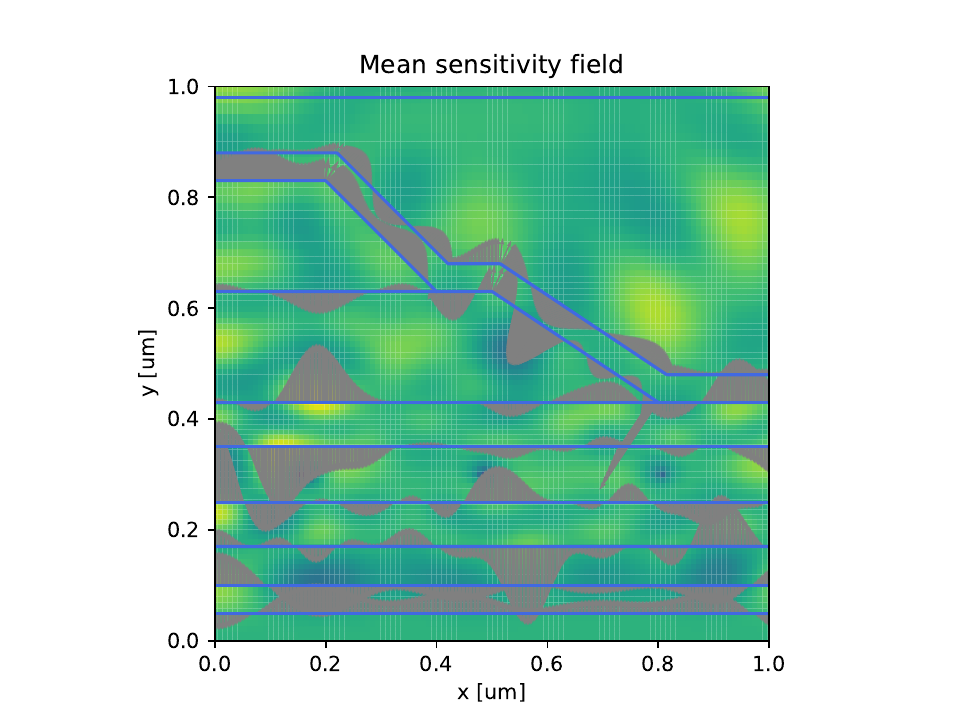}
        \caption{\label{appendix:fig:dCG_sensitivity_field_zoomed_iter0}}
    \end{subfigure}
    \hfill
    \begin{subfigure}[ht]{.47\textwidth}
        \centering
        \includegraphics[height = 9.5cm, trim = 3.4cm 0.cm 2.8cm 0cm, clip]{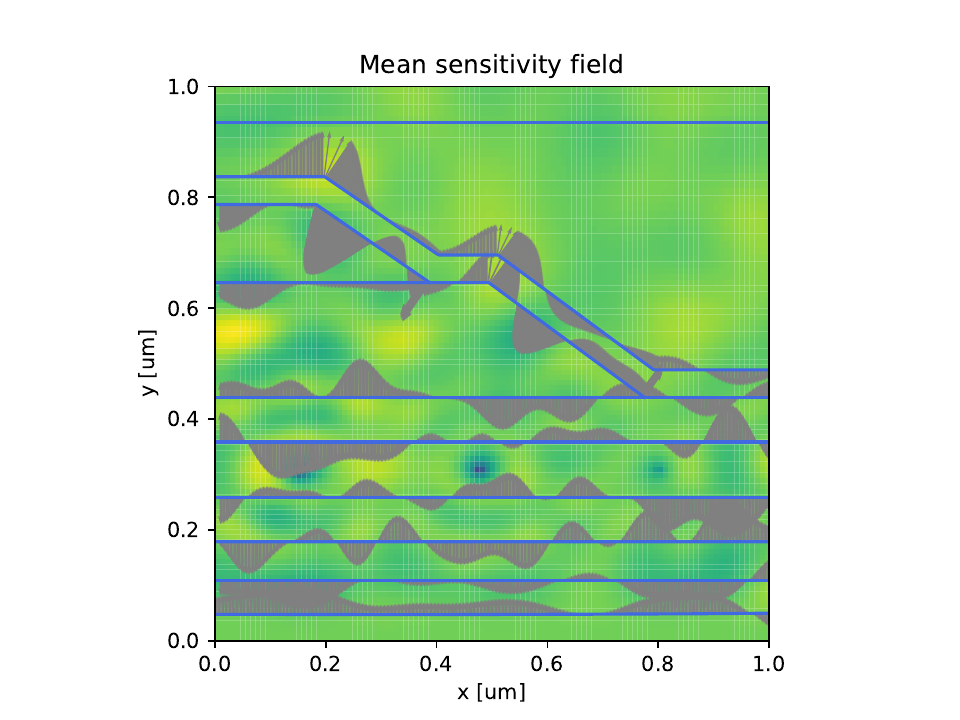}
        \caption{\label{appendix:fig:dCG_sensitivity_field_zoomed_iter29}}
    \end{subfigure}
    \caption{Comparison of the layout of the internal structure of the toy \textmu LED model. During the optimization, the optimizer proceeds to reduce the initial parameters, in particular thicknesses shown in \autoref{appendix:fig:dCG_sensitivity_field_zoomed_iter0} of multiple layers. Finally, the overall thickness of the device is reduced as shown in \autoref{appendix:fig:dCG_sensitivity_field_zoomed_iter29}. }
\end{figure*}

\subsubsection{Target}
We're interested in the enhancement of the \textmu LED emission efficiency into the forward direction. Hence, we measure the so called light extraction efficiency (LEE) into a particular solid angle $\Gamma$ which is given by the integrated power of the emission in the farfield $P(\theta, \lambda, \text{dipole}) = \frac{|E(\theta, \lambda, \text{dipole})|^2\, n\, c\, \varepsilon_0}{2}$, normalized by the injected power $P_0$ to compensate for the arbitrary power selected for the numerical computation. The constants to compute the power are the refractive index $n$ which is $n\approx 1$ for air, the speed of light $c$ and the vacuum permittivity $\varepsilon_0$. The quantities which are of interest are the LEE density $\varrho_\text{LEE}(\theta) = |\langle E(\theta)\rangle_{\lambda,\, \text{dipoles}}|^2 / P_0$ which measures how much of the injected light is emitted into a specific direction, (weighted) averaged over all wavelengths and incoherently averaged over the individual dipoles\footnote{Incoherent averaging is important for this problem. Electric and magnetic fields carry phase information which lead to constructive and destructive interference. This is very important for correlated emission such as in lasers but in \textmu LEDs the emission is spontaneous and therefore, every dipole has an individual contribution to the farfield power. Performing and incoherent average entails to first compute the power (and therefore loose the phase information) and the averaging the power emitted from any individual dipole.}. By integrating this density over the solid angle of our choice, we obtain the LEE$_{\pm\Gamma}$, the loss function for this problem $\mathcal{L} = -\int_{\pm \Gamma}\varrho_\text{LEE}(\theta)\, \d\theta $.

\subsubsection{Optimization}
The optimization starts with the geometry shown in \autoref{fig:dCG_uLED_application}. All parameters other than the material selection are up for the optimization to optimize. However, the requirement for the resulting geometry is that the ordering of the geometries cannot change and it must be possible to manufacture the geometry by a sequential process and thus, only certain types of geometrical changes are allowed. For instance, some angles and depths of geometry 1 can be changed but the following geometries must follow the contour of geometry 1, only the relative thickness can be changed. The available parameters for all geometries are shown in \autoref{tab:materials_and_parameters}. 
To optimize the geometry, we employ simple gradient descent, starting with the model shown in \autoref{fig:dCG_uLED_application}. Since the boundary gradients are hard to see in the full view of the problem, shown in \autoref{fig:dCG_uLED_optimization} we show additionally the gradients only within the \textmu LED core in \autoref{appendix:fig:dCG_sensitivity_field_zoomed_iter0} and \autoref{appendix:fig:dCG_sensitivity_field_zoomed_iter29}. The oscillations of the boundary gradients exemplify the problem encountered with the wave-optical optimization problem: moving any boundary into any direction accumulates both positive and negative contributions to the loss. Without a systematic approach at computing the sensitivities, its almost impossible to predict the impact of slight modifications of the geometrical features. 

\subsection{Constraint violation $\varepsilon$}\label{appendix:constraint_violation}

As discussed in \autoref{sec:example_opitmization}, an additional parameter $\varepsilon$ can be introduced to allow slight violations of constraints. This parameter provides the model with extra flexibility during optimization, helping it to escape local optima. Two illustrative examples of how such constraint violations can be implemented are shown in \autoref{appendix:fig:epsilon_displacement}.

In example (a), a continuous displacement of the contour of a geometry is permitted that violates the constraint that the geometry must have a perfect circular boundary. The violation is achieved through a vector field $\Vec{\varepsilon}(x,y)$ which introduces a point-wise displacement of the boundary. In contrast, example (b) depicts two circular dCGs with the constraint that their boundaries must touch. Here, a violation parameter $\Vec{\varepsilon}$ is introduced, displacing the center of the child geometry to violate the constraint. As $\Vec{\varepsilon}$ decays to zero, both examples revert to a state where all constraints are satisfied.

\begin{figure}[ht] 
    \centering 
    \includegraphics[width=\columnwidth, trim=0cm 0cm 0cm 0cm, clip]{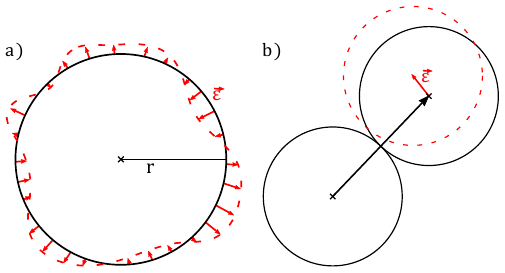} 
    \caption{
        \label{appendix:fig:epsilon_displacement} Illustrative examples of constraint violation using the parameter $\varepsilon$. (a) Continuous displacement of a geometry's contour via a vector field $\Vec{\varepsilon}(x,y)$ that violates the circular boundary constraint. (b) Displacement of the center of a child geometry to violate the constraint that boundaries must touch.
    }
\end{figure}



\subsection{Guarantees on dCG constraints - Example}\label{appendix:guarantees_on_constraints}

For the case of the circle geometry in the example, this is shown easily by investigating the gradients of the geometry definition. Consider a circle A where the optimization is supposed to be constrained such that the center of A can only be displaced on a diagonal line. The boundary of A is explicitly given by $A = \{\Vec{x} = (x, y)\, |\, x = x_0 + r\cos(\vartheta),\, y = y_0 + r\sin(\vartheta);\, \vartheta = [0, 2\pi)\}$ where $x_0$ and $y_0$ define the center position of the circle. Since the center point of $A$ moves only on a diagonal, the position is given by $x_0 = y_0 = d$. $d$ is the actual parameter determining the position of the center and gradients are backpropagated back to this initial parameter. Since the geometry update only involves $d$, the constraints are automatically obeyed. This approach can be used to restrict the geometries to other more involved constraints, too. For example, it is possible to constrain the geometry center to the boundary of its parent geometry. Since the gradients of the child can flow to its parent due to the inherent differentiability of dCGs, the relative definition of the geometries is no problem for the update of the parent because all relevant gradients from the child are backpropagated to the parent.

\end{document}